\documentclass[conference]{IEEEtran}
\IEEEoverridecommandlockouts
\usepackage{cite}
\usepackage{amsmath,amssymb,amsfonts}
\usepackage{algorithmic}
\usepackage{graphicx}
\usepackage{textcomp}
\usepackage{xcolor}
\usepackage{bm}

\usepackage{amsmath,amssymb}
\usepackage{subfigure}

\usepackage{hyperref}
\usepackage{graphicx,graphics,color,psfrag}
\usepackage{cite,balance}
\usepackage{caption}
\allowdisplaybreaks
\usepackage{algorithm}
\usepackage{algorithmic}
\usepackage{accents}
\usepackage{amsthm}

\usepackage{url}
\usepackage[english]{babel}
\usepackage{multirow}
\usepackage{enumerate}
\usepackage{cases}
\usepackage{stfloats}
\usepackage{dsfont}
\usepackage{color,soul}
\usepackage{amsfonts}
\usepackage{fancyhdr}
\usepackage{hhline}
\usepackage{array,color}
\usepackage{mathtools}
\usepackage{pifont}

\newtheorem{remark}{Remark}

\def\BibTeX{{\rm B\kern-.05em{\sc i\kern-.025em b}\kern-.08em
    T\kern-.1667em\lower.7ex\hbox{E}\kern-.125emX}}
\begin{document}

\title{Deployment Optimization for XL-IRS Assisted Multi-User Communications\\}

\author{\IEEEauthorblockN{Chao Zhou\IEEEauthorrefmark{1}, Changsheng You\IEEEauthorrefmark{1}, Tianyu Liu\IEEEauthorrefmark{1},  and Bin Lyu\IEEEauthorrefmark{2}  } 
	\IEEEauthorblockA{\IEEEauthorrefmark{1}\text{State Key Laboratory of Optical Fiber and Cable Manufacture Technology}, \text{Department of Electronic and Electrical Engineering},\\ \text{Southern University of Science and Technology}, Shenzhen, China,\\
		\IEEEauthorrefmark{2}\text{School of Communications and Information Engineering}, \text{Nanjing University of Posts and Telecommunications}, Nanjing, China}
		Emails: zhouchao2024@mail.sustech.edu.cn, youcs@sustech.edu.cn, liuty2022@mail.sustech.edu.cn, blyu@njupt.edu.cn.
		\thanks{\emph{Corresponding author: Changsheng You.}}
	}

\maketitle

\begin{abstract}
   In this paper, we study the deployment optimization for an extremely large-scale intelligent reflecting surface (XL-IRS) assisted multi-user communication system, within which the channels between the XL-IRS and the BS (or user) are modeled by the near-field spherical wavefronts. To draw some valuable insights, we first consider the single-user case, where an alternating optimization (AO) based algorithm is devised to maximize the received signal-to-noise ratio (SNR) at the user.  To address the  high computational complexity issue incurred by the AO based algorithm, three approximate received SNR expressions are obtained to yield useful insights, corresponding to the upper bound, approximate expression, and closed-form. It is demonstrated that the XL-IRS ought to be positioned near the user (rather than the BS) to obtain a higher beamforming gain. 
   Then, for the multi-user scenario, an efficient algorithm is proposed to obtain a high-quality XL-IRS placement solution by using the AO and successive convex approximation (SCA) techniques. Furthermore, the effective degree of freedom (DoF) of the BS-IRS channel is provided, which indicates that the additional effective DoF can be leveraged to improve multi-user spatial multiplexing. Last, numerical results confirm the existence of a trade-off between near-field beam-focusing gain and multiplexing gain.
\end{abstract}

\section{Introduction}

 \emph{Intelligent reflecting surface} (IRS) has emerged as a promising technology in 6G communications by dynamically controlling its reflecting elements to reconfigure the wireless propagation environment~\cite{you2024next}. This thus leads to a broad range of application scenarios, such as wireless localization, physical layer security,  mobile edge computing, and target sensing~\cite{shao2022target,WuIRS_Tut}. 

Besides the research in IRS channel estimation and beamforming, IRS deployment design also attracts growing research interests in recent years. Specifically, the authors in~\cite{you2022deploy} conducted a comparative comparison of user and base station (BS) side IRS deployment strategies, focusing on key metrics such as network coverage, channel condition, beamforming gain, and signaling overhead, and proposed a general hybrid IRS deployment strategy that effectively combines the complementary advantages of both strategies.
The authors in~\cite{WuIRS_Tut} conducted an investigation into the deployment design of IRS and its impact on the signal-to-noise ratio (SNR) of the receiver node in a point-to-point communication system. It was revealed that for maximizing the received SNR, it is advisable to place the IRS near either the access point (AP) or user. However, the deployment strategy may not hold for the multi-user scenarios. To tackle this issue, the authors in~\cite{Yuanwei_Deploy} devised the optimal deployment location of IRS  for maximizing the weighted sum-rate under three transmission schemes. While the deployment design of IRS for multi-user scenarios was addressed in~\cite{Yuanwei_Deploy}, it is not practical to consider a single-antenna transmitter. 
Therefore, the deployment strategy for an IRS-assisted multi-user system is still an open problem. 

Additionally, to enhance the reflective beamforming gain, it is desirable to deploy a massive quantity of reflecting elements at the IRS, which is referred to as extremely large-scale IRS (XL-IRS). Note that for XL-IRS systems, the user and the BS are more prone to be situated in the near-field region of the XL-IRS, due to a larger Rayleigh distance~\cite{WuPro_IEEE}. As such, the more  accurate spherical wavefronts (instead of planar wavefronts) channel model need to be considered~\cite{Wang2024}, which introduces new challenges to the XL-IRS deployment design. However, there still lacks a dedicated study to the XL-IRS deployment strategy, which thereby  motivates this work.

In this paper, we first study the impact of XL-IRS deployment location on the received SNR in a single-user scenario.
Specifically, an alternating optimization (AO) based algorithm is devised to jointly optimize the transmit beamforming vector and reflection matrix for maximizing the received SNR. 
To address the high complexity caused by the AO based algorithm, three approximate expressions for received SNR are derived, which indicate that the XL-IRS is preferable to place near the user to achieve a higher beamforming gain under the spherical wavefronts model. Then, we extend the AO based algorithm to a multi-user scenario and employ the successive convex approximation (SCA) method to address a more complex optimization problem. Moreover, the effective degree of freedom (DoF) of the BS-IRS channel is provided, which can be utilized for achieving a higher multiplexing gain. Numerical results confirm that a trade-off exists between beamforming gain and multiplexing gain.

\section{System Model and Problem Formulation}

\begin{figure}
	\centering
	\includegraphics[width=0.9\linewidth]{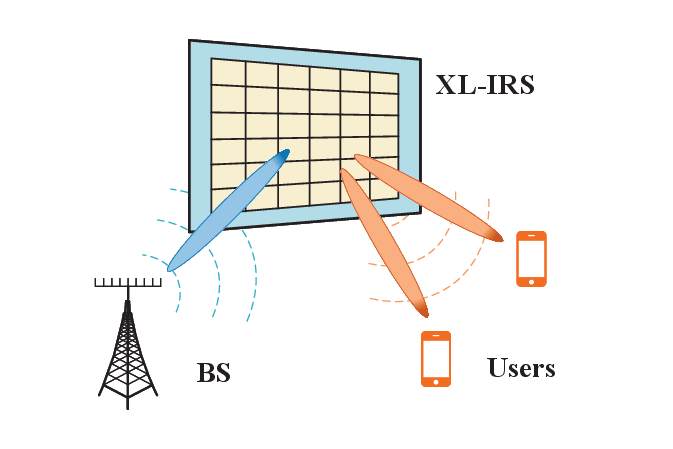}
	\caption{An XL-IRS assisted multi-user system.}
	\label{System_Model}
\end{figure}

As shown in Fig.~\ref{System_Model}, we take into account an XL-IRS assisted multi-user wireless communication system, where a uniform linear array (ULA) BS with $M$ antennas communicates with $K$ users each having a single antenna. 
We assume that the direct BS-user channels are blocked due to environmental obstacles. Thus, an $N$ reflecting elements XL-IRS is deployed for enhancing the propagation environment from the BS to each user, where $ N_x $ and $ N_y $  stand for the numbers of horizontal and vertical IRS reflecting elements, respectively.

\subsection{Channel Model}
According to~\cite{NF_SphericalWave}, the Rayleigh distance of the XL-IRS is given by $Z_{\rm B/U} = \frac{2\left(D_{\rm R}+D_{\rm B/U}\right)^2 }{\lambda}$, where $D_{\rm R} = \sqrt{  \left(N_{x}-1\right)^2 d^2   + \left( N_{y}-1\right)^2 d^2 }$, $D_{\rm B} = \left( M-1\right)d $, and $D_{\rm U}~\approx0$ are the apertures of the XL-IRS, BS and users, respectively, $d=\frac{\lambda}{2}$ represents the antenna spacing with $\lambda$ denoting the carrier wavelength. 
As the aperture of XL-IRS is sufficiently large, both the BS and users are located in the near-field region of the XL-IRS. To accurately model the propagation environment between the XL-IRS and BS (users), the general uniform spherical wavefronts is taken into account~\cite{NF_SphericalWave}.
To draw some   valuable insights into the deployment strategy of XL-IRS, the line-of-sight (LoS) channel model is assumed.\footnote{In this work, we focus on the LoS channel for simplicity. However, the proposed AO based algorithm and the closed-form in~\eqref{closed} are also applicable to multi-path scenarios.} 
Accordingly, the BS-IRS channel, denoted as $\bm G$, can be modeled as 
\begin{align}
	\left[ {\bm {G}}\right]_{n,m} = \frac{\lambda}{4\pi d_{\rm BI}} e^{j\frac{2\pi}{\lambda}d_{n,m}},
\end{align}
where $ \left[ {\bm {G}}\right]_{n,m}  $ and   $d_{n,m}$ denote the complex channel and transmission distance  from the $m$-th BS antenna to the $n$-th XL-IRS reflecting element, $d_{\rm BI}$ represents the distance between the center reference point of BS and XL-RIS.
As such, the channel between the XL-IRS and the $k$-th user, denoted as $\bm r_{k}$, is given by
\begin{align}
\left[ {\bm r}_{k}\right]_{n} = \frac{\lambda}{4\pi d_{{\rm I},k}}e^{j\frac{2\pi}{\lambda}d_{n,k}},
\end{align}
where  $ \left[ {\bm {r}}\right]_{n}  $ and  $d_{n}^{\left(k\right)} $ denote the complex channel and transmission distance from the $n$-th  XL-IRS reflecting element to the $k$-th user, $d_{{\rm I},k}$ represents the distance  between the center reference point of XL-IRS and $k$-th user.

\subsection{Signal  Model}
Let $ w_k $ denotes the transmit beamforming vector at the BS for user $k$. 
Then the signals transmitted by the BS can be represented as $\bm x = \sum_{k=1}^{K} \bm w_{k} s_{k}$,
where $s_{k}$ represents the transmitted signal for the $k$-th user with $s_{k}\sim\mathcal{CN}\left( 0,1\right) $. Note that the total transmit power at the BS  should satisfy 
\begin{align}\label{Power_constraint}
	\sum_{k=1}^{K}\left|\left|\bm w_{k} \right|  \right|^2 \le P,
\end{align}
where $P$ represents the maximum transmit power.   Then, the signal received by the $k$-th user is expressed as 
\begin{align}
	y_{k} = {\bm r}_{k}^{H} {\bm \Theta} {\bm G} \sum_{k=1}^{K} \bm w_{k} s_{k} + z_{k}, 
\end{align}
where $ \bm \Theta =\text{diag}\left(\phi_{1},\phi_{2},\ldots,\phi_{N} \right) $ represents the reflection matrix with $| \phi_{n}| = 1$ and $z_{k} \sim \mathcal{CN}\left(0,\sigma_k^2 \right)$ with $\sigma_k^2$ denoting the noise power. 
As such, the sum-rate in bits/second/Hertz (bit/s/Hz), denoted as $R_{\rm sum}$, is given by
\begin{align}
	R_{\rm sum} = \sum_{k=1}^{K}\log_2\left(1+ \frac{ \left| {\bm r}_{k}^{H} {\bm \Theta } {\bm G} \bm w_{k} \right| ^2}
	{\sum_{j\neq k}^{K}\left| {\bm r}_{k}^{H} {\bm \Theta  } {\bm G} \bm w_{j} \right| ^2 + \sigma_{k}^2  } \right).
\end{align}
\subsection{Problem Formulation}
We aim to investigate the impact of XL-IRS deployment on system performance, taking into account the near-field uniform spherical wavefronts. To achieve this, we have formulated an optimization problem as follow for maximizing the sum-rate of all users by jointly designing the transmit beamforming and reflection matrix. 
\begin{align}
	\textbf{(P1)}:~~& \max_{\bm w_k, \bm \Theta}~~R_{\rm sum} \nonumber \\
	\textrm{s.t.}&~~\eqref{Power_constraint}, \nonumber\\
	&~{\rm arg}(\phi_{n}) \in \left[ 0,2\pi\right) ,~n=1,2,\ldots,N.\label{Eq_phase}  
\end{align}
where~\eqref{Power_constraint} is the transmit power constraint and~\eqref{Eq_phase} is the phase shift  constraint.
	
\section{Signal-User Scenario}
\begin{figure*}[t]
	\centering
	\subfigure[$M=1$.]{\label{M1_SU}
		\includegraphics[height=4.2cm]{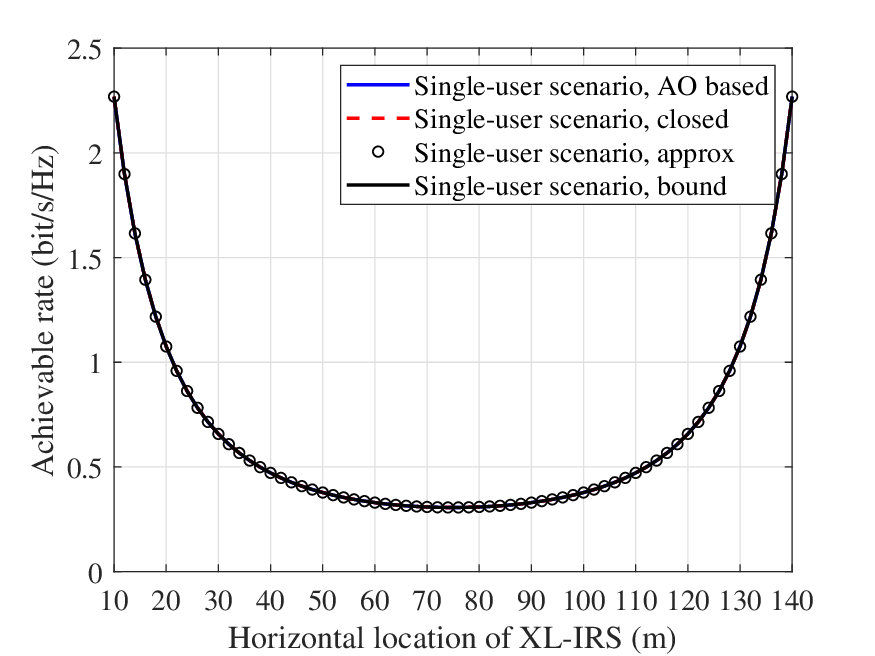}}
	\hspace{5pt}
	\subfigure[$M=16$.]{\label{M16_SU}
		\includegraphics[height=4.2cm]{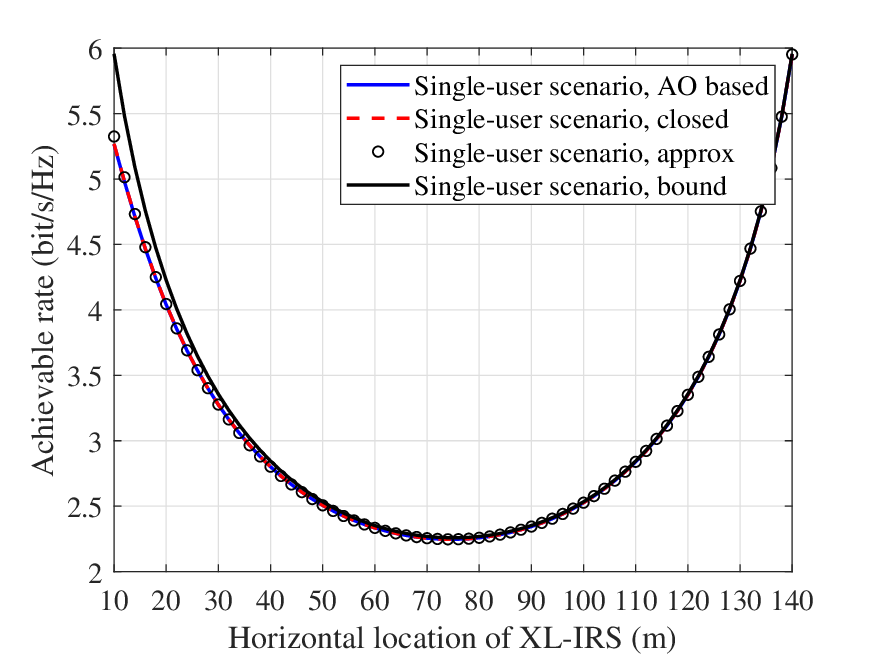}}
	\hspace{5pt}
	\subfigure[$M=64$.]{\label{M32_SU}
		\includegraphics[height=4.2cm]{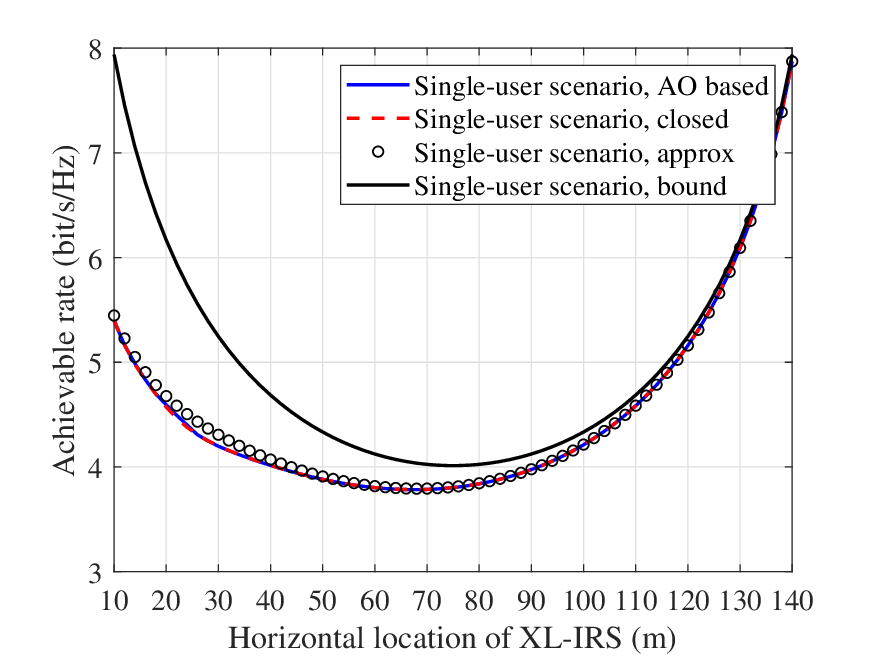}}
	\caption{Achievable rate under the single-user scenario with different transmit antennas, $M$}
	\label{SU_approx}
	\vspace{-5pt}
\end{figure*}

We first consider the signal-user scenario (i.e., $ K=1 $) to obtain valuable insights. Under this setup, there is no interference between users. As such, problem $ \textbf{(P1)} $  can be simplified to
\begin{align}
		\textbf{(P2)}:~~& \max_{\bm w_1, \bm \Theta}~~  \left| {\bm r}_{1}^{H} {\bm \Theta } {\bm G} \bm w_{1} \right|^2 \nonumber \\
		\textrm{s.t.}
		&~~\eqref{Eq_phase},  \nonumber \\
		&~~\left| \left| \bm w_1\right| \right|^2 \le P.
\end{align}
Note that \textbf{(P2)} is a non-convex optimization problem due to the coupling variables (i.e., $\bm w_1$ and $\bm \Theta$) in the objective function. Although semi-definite relaxation (SDR) technique can be applied to obtain a high-quality solution to this problem,  the computational complexity is significantly high, especially when optimizing the reflection matrix. To tackle this challenge, an AO based algorithm is designed for addressing this problem~\cite{WuQINGQing2019}. Specifically, for a given reflection matrix $\bm \Theta$, the optimal transmit beamforming vector is obtained by 
\begin{align}\label{optimal_w_su}
	\bm w_1 = \sqrt{P}\frac{\left({\bm r}_{1}^{H} {\bm \Theta } {\bm G}  \right)^H }{|| {\bm r}_{1}^{H} {\bm \Theta } {\bm G}||  }. 
\end{align}
Next, given any feasible transmit beamforming vector $\bm w_{1}$, the optimal phase shift of the IRS can also be obtained, the $n$-th element of which satisfies
\begin{align}
	\phi_{n}^{*} = -{\rm arg}\left( \text{diag}\left( \bm r_{1}^H\right)\bm G \bm w_{1}  \right). 
\end{align}
By iteratively optimizing the transmit beamforming vector and phase shifts, the optimal solution to \textbf{(P1)} can be found.
 While the complexity of AO based algorithm is related to the number of $N$, which is significantly high when deploying an XL-IRS. To address this issue, we then analyze the structure of the objective value in \textbf{(P2)} for presenting a closed-form solution. Specifically, by substituting~\eqref{optimal_w_su} into problem \textbf{(P2)}, we can simplify the problem  as
\begin{align}
	\textbf{(P2.1)}:~~& \max_{\bm \Theta}~~ \left| \left|{\bm r}_{1}^{H} {\bm \Theta } {\bm G} \right| \right|^2 \nonumber \\
	\textrm{s.t.}
	&~~\eqref{Eq_phase}.  \nonumber 
\end{align} 
\underline{{$ \bm {M=1}$}:}
To obtain some useful insights, we first consider that the transmit antenna $M=1$, and the BS-IRS channel degenerates into $\bm g=\bm G(:,1)$. Under this setup, the optimal phase shift of the $n$-th reflecting element is $\phi_{n}^{*}=e^{-j\frac{2\pi}{\lambda}(d_{n,1}+d_{n}^{\left(1 \right) })}$. It should be noted that the first term is intended to offset the phase difference in $\bm g$, while the second term is intended to align with $\bm r_{1}$  to obtain the maximum beamforming gain at the user. Therefore, the received SNR is only related to the location of XL-IRS. Let assume that the distance between the XL-IRS and user is constant, and the XL-IRS is deployed between them. Thus, the received SNR is simplified as
\begin{align}
	{\rm SNR}_{1}=\frac{\beta N^2}{d_{\rm BI}^2 d_{{\rm I},1}^2\sigma_{1}^2},
\end{align}
where $\beta=\left( \frac{\lambda}{4\pi}\right)^4$. It is evident that placing the XL-IRS near the BS or user is able to achieve the highest SNR. \\
\underline{{$\bm  {M>1}$}:}
Subsequently, we proceed to analyze the scenario when $M >1$. Assuming that the phase differences in $\bm r_{1}$ and $\bm G$  is perfectly offset, the upper bound of the received SNR can be achieved, which is given by
\begin{align}\label{UB}
	{\rm SNR}_{1}^{\rm bound}=\frac{\beta MN^2}{d_{\rm BI}^2 d_{{\rm I},1}^2\sigma_{1}^2}.
\end{align}
Due to the spherical wavefronts, the expression~\eqref{UB} is inaccurate. We then analyze the objective value in problem \textbf{(P2)}.
Given that the phase difference in $\bm r_{1}$ can be offset by adjusting the phase of each XL-IRS element, the optimization problem \textbf{(P2.1)} can thus be reformulated as
\begin{align}
	\textbf{(P2.2)}:~~& \max_{\bm \theta}~~ \left| \left| {\bm \theta }^H {\bm G} \right| \right|^2 \nonumber \\
	\textrm{s.t.}
	&~~\eqref{Eq_phase},  \nonumber 
\end{align}
where $\bm \theta =\left[\phi_{1},\phi_{2},\ldots,\phi_{N} \right]^T  $. Let  $\mu_{i}$ denote the $i$-th largest  eigenvalue to the matrix $\bar {\bm  G}= \bm G \bm G^H$, and $\bm \psi_i$ represent its corresponding unit eigenvector.
As we known, the optimal solution to \textbf{(P2.2)} is $\bm \theta = \sqrt{N} \bm \psi_1$ when ignoring the constant modulus constraint~\eqref{Eq_phase}. Thus, we can obtain an approximate expression for the received SNR when $M>1$, which is expressed as 
\begin{align}\label{approx}
	 {\rm SNR}_{1}^{\rm approx}=\frac{\beta N \mu_{1}}{d_{\rm BI}^2 d_{{\rm I},1}^2\sigma_{1}^2}.
\end{align} 
While in most cases the deviation between the approximate expression and the theoretical optimal value (i.e., the value obtained by AO based algorithm) can be disregarded, as confirmed in Fig.~\ref{SU_approx}, there is still a noticeable discrepancy when $M$ increases. To address this issue, we take the constant modulus constraint~\eqref{Eq_phase} into consideration to obtain an accurate closed-form.
As such,  we have $\bm \theta_i = e^{\text{arg}\left( \bm \psi_i \right) }$. Let denote  $\kappa_{i} =  \frac{\bm \theta_i^H \bar {\bm  G} \bm \theta_i}{\psi_i^H \bar {\bm  G} \psi_i}  $ as correlation ratio. Then,  problem \textbf{(P2.2)} can be rewritten as 
\begin{align}
	\textbf{(P2.3)}:~~& \max_{i}~~ \frac{\bm \theta_{i}^H \bar {\bm  G} \bm \theta_{i}}{\psi_i^H \bar {\bm  G} \psi_i} \psi_i^H \bar {\bm  G} \psi_i,
	 \nonumber 
\end{align}
which means that the correlation ratio (i.e., $\kappa_{i}$) and the $i$-th largest eigenvalue (i.e., $ \psi_i^H \bar {\bm  G} \psi_i $) are both large enough to achieve the highest received SNR, which  is expressed as
\begin{align}\label{closed}
	{\rm SNR}_1^{\rm closed} = \frac{\beta N  \max_{i}{\{ \kappa_{i}\mu_{i}\}}  }{d_{\rm BI}^2 d_{{\rm I},1}^2\sigma_{1}^2},
\end{align}
where $ \kappa_{i}\mu_{i}$ represents the beamforming gain.

\begin{figure}[b]
	\centering
	\includegraphics[width=0.9\linewidth]{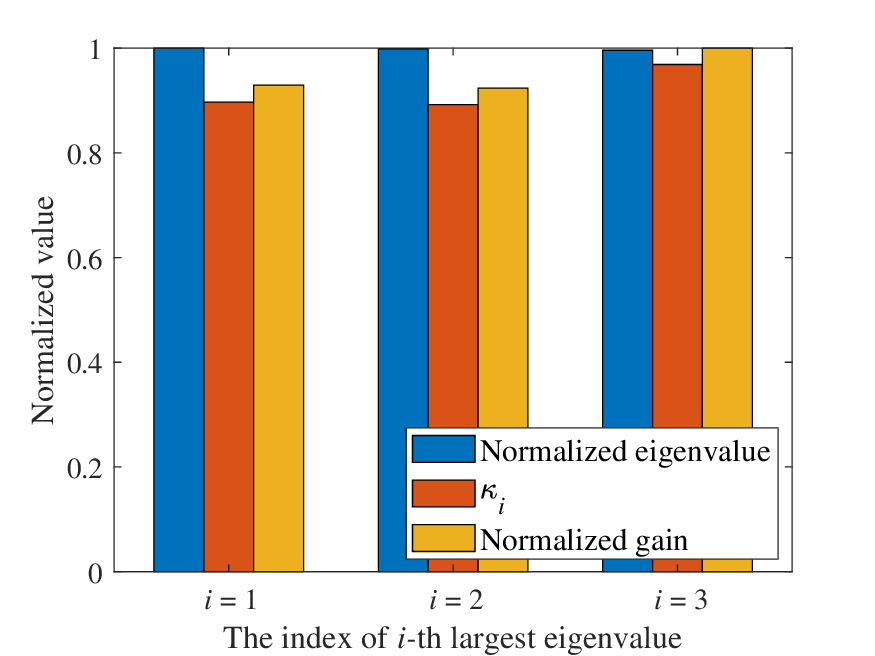}
	\caption{Normalized value.}
	\label{Eigenvalue}
\end{figure}
\begin{remark}[Effect of spherical wavefronts on received SNR~\footnote{Note that the received SNR expressions are obtained under the assumption of uniform spherical wavefronts, thus~\eqref{UB},~\eqref{approx}, and~\eqref{closed} may be inaccurate as $N \to \infty$.}]
	\rm Considering that the phase difference in $\bm r_1$ can always be offset, the received SNR is primarily determined by the multiple-input multiple-output (MIMO) channel from the BS to the XL-IRS (i.e., $\bm G$). When the XL-IRS is located near the BS, the rank of $\bm G$ is typically greater than $1$ due to spherical wavefronts, making it difficult to perfectly offset the phase difference in $\bm G$ and hence resulting in significant disparity between~\eqref{UB} and~\eqref{closed}. However, when the XL-IRS is deployed near the user, spherical wavefronts gradually transition to planar wavefronts and  $\text{rank}(\bm G) \approx1$~\cite{Liu_Tut}. Consequently, it becomes possible to offset  phase difference in $\bm G$, leading to only a marginal gap between~\eqref{UB} and~\eqref{closed}.
\end{remark}

\begin{remark}[Low complexity and high accuracy solution]
	\rm In~\eqref{closed}, we present a closed-form expression for the received SNR in a single-user scenario, without the need for 
	iteratively optimizing the transmit beamforming and reflection matrix. This approach effectively \emph{reduces computational complexity}. Furthermore, our derived closed-form \emph{achieves higher accuracy} than the upper bound (i.e.,~\eqref{UB}) and approximate expression (i.e.,~\eqref{approx}), which is very close to the result obtained by the AO based algorithm. Moreover,~\eqref{closed} indicates that the received SNR is related not only to the eigenvalues, but also to the correlation ratio. Thus, for the scenario that $\text{rank}\left(\bar{\bm  G} \right)>1 $, $i$ may not be set as $1$. As shown in Fig.~\ref{Eigenvalue}, although $\mu_{1}$ is the largest eigenvalue, the normalized gain is higher when $i=3$ because of a higher correlation ratio $\kappa_{3}$.
\end{remark}

\begin{remark}[Deployment strategy in single-user scenario] 
	\rm The simulation results provided in Fig.~\ref{SU_approx} demonstrate that for $M=1$, the XL-IRS can be deployed at both the BS side and user side. However, as the increase of transmit antennas, it is more advantageous to deploy the XL-IRS at the user side. This discrepancy arises from the spherical wavefronts, which makes it difficult  to offset the phase difference in $\bm G$ when the XL-IRS is located close to the BS. 
	Conversely, when the XL-IRS is deployed near the user, it allows for an approximate modeling of the BS-IRS channel $\bm G$ as a far-field channel with a rank of $1$. As a result, this enables perfect offsetting of phase differences in $\bm G$, leading to a near-optimal achievable rate (i.e., upper bound in Fig.~\ref{M32_SU}).
\end{remark}

\section{Multi-User Scenario}
We study in this section the deployment strategy in a multi-user scenario. Specifically, we first propose an AO based algorithm for maximizing the sum-rate through alternately optimizing the transmit vector $\bm w_{k}$ and reflection matrix $\bm \Theta$.
\subsection{Optimization of transmit beamforming}
 We first design the transmit beamforming vector $\bm w_{k}$ given fixed $\bm \Theta$. The objective value is non-convex due to the complex signal-to-interference-plus-noise-ratio (SINR) and the quadratic form expressions.
To address this issue, we employ the SCA technique to approximate \textbf{(P1)} by a convex optimization problem~\cite{zhou2023}, which is given by
\begin{align}
			&R_{k} = \log_2\left(1+ \frac{ \left| {\bm r}_{k}^{H} {\bm \Theta } {\bm G} \bm w_{k} \right| ^2}
		{\sum_{j\neq k}^{K}\left| {\bm r}_{k}^{H} {\bm \Theta  } {\bm G} \bm w_{j} \right| ^2 + \sigma_{k}^2  } \right) \nonumber \\
		&\ge \frac{1}{\ln2}\Bigg(  
		\ln  \bigg( 1+\frac{| {\hat{a}}_{k}^{\left(t \right) } |^2}{  {\hat{b}}_{k}^{\left(t \right) }  } \bigg)     
		-   \frac{| {\hat{a}}_{k}^{\left(t \right) } |^2}{  {\hat{b}}_{k}^{\left(t \right) }  } 
		+ \frac{2\mathcal{R}\left\lbrace\left(  {\hat{a}}_{k}^{\left(t \right) } \right)^H  {\hat{a}}_{k}  \right\rbrace }{{\hat{b}}_{k}^{\left(t \right) }}\nonumber \\
		&-\frac{| {\hat{a}}_{k}^{\left(t \right) } |^2 \big(  | {\hat{a}}_{k} |^2  + {\hat{b}}_{k}  \big)  }
		{ {\hat{b}}_{k}^{\left(t \right) } \big(| {\hat{a}}_{k}^{\left(t \right) } |^2+   {\hat{b}}_{k}^{\left(t \right) }\big) }
		\Bigg) \triangleq {\hat{R}}_{k},
\end{align}
where ${\hat{a}}_{k} =  {\bm r}_{k}^{H} {\bm \Theta } {\bm G} \bm w_{k}$,  ${\hat{b}}_{k} =  \sum_{j\neq k}^{K}\left| {\bm r}_{k}^{H} {\bm \Theta  } {\bm G} \bm w_{j} \right| ^2 + \sigma_{k}^2 $, ${\hat{a}}_{k}^{\left(t \right) }$ and ${\hat{b}}_{k}^{\left(t \right) }$ are the feasible points of ${\hat{a}}_{k}$ and ${\hat{b}}_{k}$ in the $t$-th iteration. Thus, the  problem related to transmit beamforming optimization can be rewritten as 
\begin{align}
	\textbf{(P3.1)}:~~& \max_{\bm w_k}~~\hat{R}_{\rm sum} \nonumber \\
	\textrm{s.t.}&~~\eqref{Power_constraint}, \nonumber
\end{align}
where  $ \hat{R}_{\rm sum} = \sum_{k=1}^{K}{\hat{R}}_{k} $. Since  problem \textbf{(P3.1)} is convex, which  can be  directly solved using the CVX tool~\cite{grant2014cvx}.

\subsection{Optimization of reflection matrix}
After obtaining the $\bm w_{k}$, we then optimize the reflection matrix $\bm \Theta$. By using ${\bm r}_{k}^{H} {\bm \Theta } {\bm G} \bm w_{k} = {\bm r}_{k}^{H} \text{diag}\left({\bm G} \bm w_{k} \right)  \bm \theta $, the SCA method can also be utilized to reformulate the optimization problem. 
Similarly, the lower-bound of achievable rate for the $k$-th user can be expressed as
\begin{align}
	& R_{k} = \log_2\left(1 + \frac{ \left|{\bm r}_{k}^{H} \text{diag}\left({\bm G} \bm w_{k} \right)  \bm \theta \right|^2 }{\sum_{j\neq k}^{K} \left|{\bm r}_{k}^{H} \text{diag}\left({\bm G} \bm w_{j} \right)  \bm \theta \right|^2  +\sigma_{k}^2}  \right)  \nonumber\\
	&\ge \frac{1}{\ln2}\Bigg(  
	\ln  \bigg( 1+\frac{| {\tilde{a}}_{k}^{\left(t \right) } |^2}{  {\tilde{b}}_{k}^{\left(t \right) }  } \bigg)     
	-   \frac{| {\tilde{a}}_{k}^{\left(t \right) } |^2}{  {\tilde{b}}_{k}^{\left(t \right) }  } 
	+ \frac{2\mathcal{R}\left\lbrace\left(  {\tilde{a}}_{k}^{\left(t \right) } \right)^H  {\tilde{a}}_{k}  \right\rbrace }{{\tilde{b}}_{k}^{\left(t \right) }}\nonumber \\
	&-\frac{| {\tilde{a}}_{k}^{\left(t \right) } |^2 \big(  | {\tilde{a}}_{k} |^2  + {\tilde{b}}_{k}  \big)  }
	{ {\tilde{b}}_{k}^{\left(t \right) } \big(| {\tilde{a}}_{k}^{\left(t \right) } |^2+   {\tilde{b}}_{k}^{\left(t \right) }\big) }
	\Bigg) \triangleq {\tilde{R}}_{k},
\end{align}
where  ${\tilde{b}}_{k}=\sum_{j\neq k}^{K} \left|{\bm r}_{k}^{H} \text{diag}\left({\bm G} \bm w_{j} \right)  \bm \theta \right|^2 + \sigma_{k}^2 $, ${\tilde{a}}_{k} =  {\bm r}_{k}^{H} \text{diag} \left( {\bm G} \bm w_{k}  \right) \bm \theta  $, ${\tilde{a}}_{k}^{\left(t \right) }$ and ${\tilde{b}}_{k}^{\left(t \right) }$ are the feasible points of ${\tilde{a}}_{k}$ and ${\tilde{b}}_{k}$ in the $t$-th iteration.  
Accordingly, the optimization problem of optimizing the reflection vector $\bm \theta$ can be reformulated as 
\begin{align}
	\textbf{(P3.2)}:~~& \max_{\bm \theta}~~\tilde{R}_{\rm sum} \nonumber \\
	\textrm{s.t.}&~~|\phi_{n}|^2 = 1,~n=1,2,\ldots,N,
\end{align}
where $ \tilde{R}_{\rm sum} = \sum_{k=1}^{K}{\tilde{R}}_{k} $.
Since problem \textbf{(P3.2)} is convex, the CVX tool is utilized to solve it. Note that the convergence of the proposed AO-based algorithm can be guaranteed, and the computational complexity can be referred to in~\cite{zhou2023}. Compared to the single-user scenario,  optimization processes of the transmit beamforming vector and reflection matrix are more complex, while some insights can be obtained.

Based on the results for the single-user scenario, the XL-IRS demonstrates strategic deployment near the base station or user in order to mitigate path loss. Furthermore, as the increase of   transmit antennas, the XL-IRS exhibits a stronger inclination towards the user side. However, for a multi-user case, the optimal XL-IRS deployment strategy is more complicated. We denote the MIMO channel from the BS to all the users as
\begin{align}
	\bm H =\left[ {\bm r}_{1}^{H} {\bm \Theta } {\bm G};\ldots;{\bm r}_{K}^{H} {\bm \Theta } {\bm G}  \right]=\left[{\bm r}_{1}^{H};\ldots;{\bm r}_{K}^{H} \right] {\bm \Theta } {\bm G}, 
\end{align} 
whose rank is less than $\min\{K,M\}$. We assume that the channel from the XL-IRS to each user (i.e., $\bm r_{k}$) is independent, and the rank of $\bm H =  \min\{K,\text{rank}\left(\bm G \right) \}$. Let $\rm EDoF$  denotes the effective DoF~\cite{NF_SphericalWave},  which is defined as
\begin{align}
	&\text{EDoF} = \left( \frac{\text{tr}\left( \bar{\bm  G}\right)    }{||\bar{\bm  G} ||_F}\right) ^2.
\end{align} 
For the multi-user scenario, at least $K$ independent channels are required to support independent data streams for all $K$ users. In other words, 
$ \text{EDoF}\left(\bm G \right) \ge K$ should be satisfied.

\begin{figure}[h]
	\centering
	\includegraphics[width=0.8\linewidth]{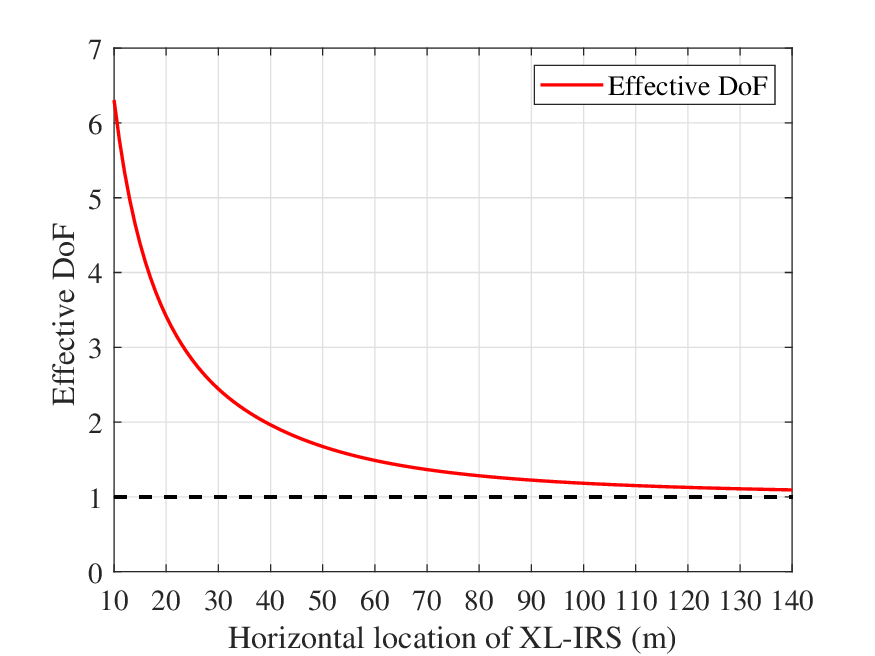}
	\caption{DoF versus horizontal location of XL-IRS.}
	\label{DoF}
\end{figure}
\begin{remark}[DoF analysis]
	\rm In Fig.~\ref{DoF}, the effective DoF of the channel $\bm G$ with respect to the horizontal location of XL-IRS is provided. As depicted in Fig.~\ref{DoF}, it is observed that the effective DoF decreases as the distance increases, indicating that spherical wavefronts introduce additional DoF for near-field LoS channel compared to planar wavefronts. Moreover, when the horizontal location of XL-IRS increases, the effective DoF gradually approaches to one, suggesting that concentrating all transmit power on a single user yields a better performance.
\end{remark}

\begin{remark}[Trade-off between beamforming gain and multiplexing gain]
	\rm For the multi-user scenario, the deployment of XL-IRS demonstrates distinct characteristics. Specifically, its placement at the BS side provides an increased DoF for spatial multiplexing, while its placement at the user side offers a higher near-field beam-focusing gain with a trade-off in effective DoF. This observation underscores the importance of tailoring XL-IRS deployment to specific scenarios; for applications requiring high beamforming gain for a specific user, placement at the user side is preferable, whereas for scenarios necessitating high spatial multiplexing, placing it near the BS is more advantageous.
\end{remark}

\section{Numerical Results}\label{section-IV}

\begin{table}[b]
	\renewcommand\arraystretch{1}
	\centering
	\caption{Simulation Parameters}
	\label{SimulationParameters}
	\begin{tabular}{c|c}
		\hline 
		\textbf{Parameter} & \textbf{Value}          
		\\ \hline Carrier wavelength, $\lambda$ & $0.03$ m
		\\ \hline Number of antennas at the BS, $M$ & $64$ 
		\\ \hline Number of reflecting elements at the XL-IRS, $N$ & $480$ 
		\\ \hline Number of users, $K$, & 2
		\\ \hline Transmit power at the BS, $P$  & $30$  dBm
		\\ \hline Noise power, $\sigma_{k}^2$,  &  $-90$ dBm
		\\ \hline Rayleigh distance, $Z_{\rm B}$ and $Z_{\rm U}$,  & $272$  m and $213$ m
		\\ \hline Location of BS, & $ \left(0,0,0\right) $  m
		\\ \hline Location of XL-IRS, & $ \left(0,x_{\rm I},0\right) $  m
		\\ \hline Location of user $1$, & $ \left(0,150,0\right) $  m
		\\ \hline
	\end{tabular}
\end{table}
We provide simulation results in this section to validate the deployment strategy of the XL-IRS assisted wireless communication systems, and the simulation parameters are set as in Table~\ref{SimulationParameters}. Besides, the other $K-1$ users are randomly located on a circle centered at $ \left(0,150,0\right) $ with a radius of 5 meters.

\begin{figure}[h]
	\centering
	\includegraphics[width=0.8\linewidth]{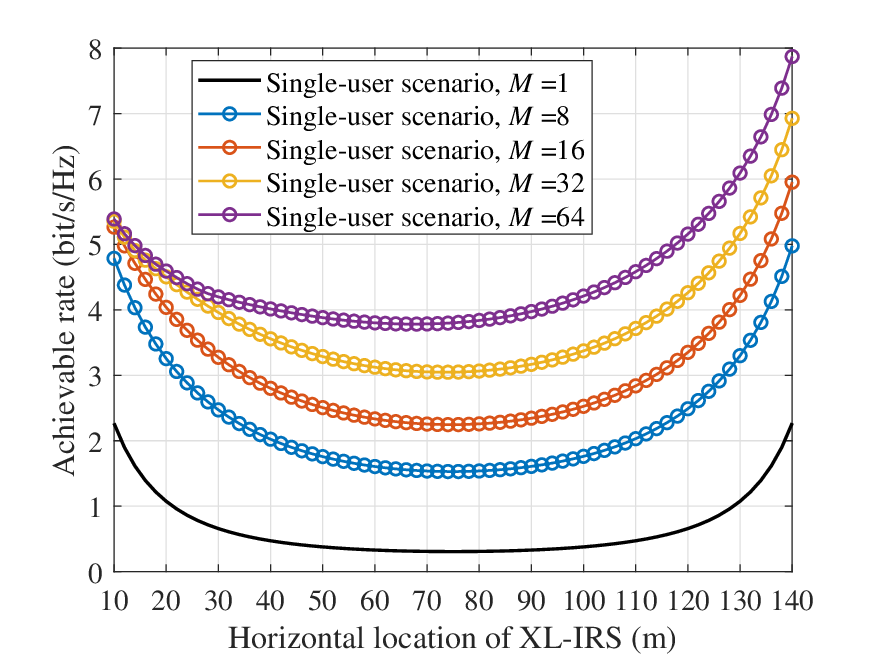}
	\caption{Achievable rate under the single-user scenario with different transmit antennas.}
	\label{SU_achieveRate}
\end{figure}

In Fig.~\ref{SU_achieveRate}, we plot the achievable rate under the single-user scenario with respect to the horizontal location of XL-IRS (i.e., $x_{\rm I}$) by different transmit antennas. Several important observations can be made. First, the deployment of XL-IRS tends to be either on the BS side or the user side to minimize the product-distance path-loss. Second, compared to the XL-IRS placement at the BS side, the user side deployment leads to superior achievable rates, particularly when $M\gg 1$ (e.g., 7.9 bit/s/Hz versus 5.4 bit/s/Hz for $M=64$). Furthermore, marginal performance improvement is achieved with an increase in the number of transmit antennas when the XL-IRS is placed at the BS side (e.g., 5.39 bit/s/Hz for $M=64$ versus 5.35 bit/s/Hz for $M=32$), attributed to imperfect offsetting of phase differences in the BS-IRS channel.

\begin{figure}[h]
	\centering
	\includegraphics[width=0.8\linewidth]{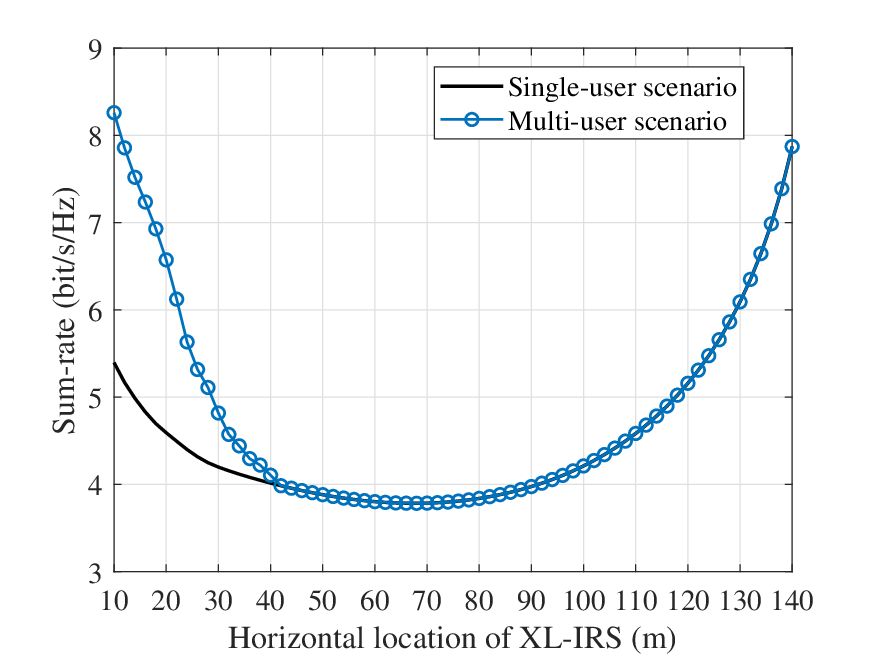}
	\caption{Achievable sum-rate under the multi-user scenario.}
	\label{MU_sumrate}
\end{figure}

Fig.~\ref{MU_sumrate} illustrates the achievable sum-rate of the multi-user scenario versus the horizontal location of XL-IRS. A notable observation is that, when XL-IRS is placed near the BS, the sum-rate surpasses that of the single-user scenario by a significant margin, potentially even doubling it. This can be attributed to the increased DoF in the BS-IRS channel, which provides a  higher spatial multiplexing gain. When $x_{\rm I}$ exceeds $42$ m, the sum-rate equals the achievable rate of the single-user scenario. This indicates that all transmit power is allocated to enhance beamforming gain for a specific user, rather than for spatial multiplexing. Therefore, a trade-off exists between the beamforming gain and multiplexing gain in XL-IRS assisted wireless communication systems, and the XL-IRS is placed either at the BS side for spatial multiplexing or the user side for beam-focusing.

\section{Conclusions}
 
In this paper, we studied the deployment strategy of XL-IRS assisted multi-user systems by taking into account spherical wavefronts. Specifically, for the  single-user scenario, an AO based algorithm and three low-complexity received SNR expressions were presented, which imply that the XL-IRS is preferred to place near the user for simultaneously reducing the product-distance path-loss and achieving a higher beamforming gain. We then extended the AO based algorithm to the multi-user scenario by solving the complex variable coupling optimization problem. Furthermore, we analyzed the effective DoF of the BS-IRS channel. It was revealed  that placing the XL-IRS near the BS can provide additional DoF for spatial multiplexing, leading to a higher  sum-rate. While for the purpose of achieving a higher beamforming gain, it is better to deploy the XL-IRS near the user. Finally, numerical results also confirmed this observation.

\section*{Acknowledgment}

This work was supported in part by the National Natural Science Foundation of China under Grant 62201242, 62331023, in part by the National Key R\&D Program Youth Scientist Project under Grant 2023YFB2905100, in part by Natural Science Foundation of Guangdong Province under Grant 2024A1515010097, in part by Shenzhen Science and Technology Program under Grant 20231115131633001, in part by the High Level of Special Funds under Grant G030230001, and in part by the Young Elite Scientists Sponsorship Program by Jiangsu Province under Grant JSTJ-2024-417.

\bibliographystyle{IEEEtran}
\bibliography{v1.bib}

\end{document}